# Low Temperature Crystal Structures and Magnetic Properties of the V$_4$-Cluster Compounds Ga$_{1-x}$Ge$_x$V$_4$S$_8$

Daniel Bichler, Herta Slavik and Dirk Johrendt

Department Chemie und Biochemie, Ludwig-Maximilians-Universität München, Butenandtstrasse 5–13 (Haus D), 81377 München, Germany

Reprint requests to D. Johrendt: Email johrendt@lmu.de

The solid solution Ga$_{1-x}$Ge$_x$V$_4$S$_8$ ($x$ = 0 - 1) was synthesized by solid state reactions and characterized by temperature-dependent x-ray powder diffraction and static magnetic susceptibility measurements. The compounds crystallize in the cubic GaMo$_4$S$_8$-type structure (space group $F\bar{4}3m$), built up by heterocubane-like [V$_4$S$_4$]$^{(5-x)+}$ cubes and [Ga$_{1-x}$Ge$_x$S$_4$]$^{(5-x)-}$ tetrahedra arranged in a NaCl-like manner. The successive substitution of Ga$^{3+}$ by Ge$^{4+}$ increases the electron count in the molecular orbital (MO) of the V$_4$-cluster gradually from seven to eight. We observe an almost linear increase of the magnetic moments, connected with a transition from ferromagnetic to antiferromagnetic ordering around $x \approx 0.5$. Remarkably, the low temperature structural phase transitions as known from the ternary compounds were also detected in the solid solution. The gallium-rich compounds (0 ≤ $x$ < 0.5) undergo rhombohedral distortions like GaV$_4$S$_8$ (space group $R3m$), whereas distortions to orthorhombic symmetry (space group $Imm$2) as known from GeV$_4$S$_8$ occur in the germanium-rich part of the solid solution (0.5 ≤ $x$ ≤ 1).

**Introduction**

Transition metal chalcogenides with the cubic $GaMo_4S_8$-type structure [1, 2] have been a matter of increasing interest for years because of their physical properties, among them superconductivity under pressure [3, 4], metal-insulator transition [5], $4d$-ferromagnetism [6] and various structural and magnetic instabilities at low temperatures [7-12]. All these phenomena reflect the strong coupling of structural, electronic and magnetic degrees of freedom in this system. Although the crystal structure of the compounds $AM_4Q_8$ (space group $F\bar{4}3m$; $A$ = Ga, Ge; $M$ = V, Ti, Nb, Ta; $Q$ = S, Se) [13-17] can easily be derived from the spinel-type $AM_2Q_4$ (space group $Fd\bar{3}m$) by an ordered half-occupation of the $A$ site and a shift of the $M$-site from 16$d$ (5/8,5/8,5/8) to 16$e$ ($x,x,x$), another description has turned out to be more useful [15]: As a consequence of the coordinate shift to $x \approx 0.6$, the transition metal atoms join to tetrahedral $M_4$-metal cluster units with strong metal-metal bonds. Thus the $GaMo_4S_8$-type structure of $GaV_4S_8$ can be described as a NaCl-like arrangement of $[V_4S_4]^{5+}$ cubes and $[GaS_4]^{5-}$ tetrahedra as depicted in Fig. 1.

The localization of electrons in metal-metal bonds causes non-metallic magnetic properties and thus these materials represent a special class of Mott insulators [9]. In order to explain the magnetic insulating properties, we have introduced a concept of appropriate $M_4$ cluster molecular orbitals (MO) [9, 15]. According to this, the metal centered electrons (*i. e.* those not incorporated in metal-ligand bonds) can occupy six bonding and six anti-bonding MO, thus the strongest $M-M$ bonds are expected to occur when the bonding set is completely filled with 12 electrons. This is almost the case in $GaMo_4S_8$ with 11 electrons per $Mo_4$ according to an idealized ionic formula separation $Ga^{3+}(Mo^{3.25+})_4(S^{2-})_8$, whereas the cluster MO of $GaV_4S_8$ is filled with seven electrons only as shown in Fig. 2. Since the three highest bonding MO ($t_2$-set) are 3-fold degenerated with respect to the $\bar{4}3m$ symmetry of the cluster, one electron remains unpaired and induces the magnetic properties.

The MO cluster approach has not only rationalized the magnetism, but explains also the structural instabilities which appear frequently this system. As an example, it has long been known that $GaMo_4S_8$ undergoes a structural distortion at 45 K [8], where the symmetry is reduced to $R3m$ and the $Mo_4$ cluster gets compressed along the 3-fold axis. On the other hand, also the isotypic vanadium compound $GaV_4S_8$ shows a rhombohedral distortion, but in this case the $V_4$-cluster becomes elongated [9]. The cluster-MO explains both effects by removing the degeneracy of the $t_2$-MO and stabilizing a particular structural distortion depending on the electron count.

Several studies have been consistently shown, that the structural and magnetic properties of $GaMo_4S_8$-type compounds are tunable by the variation of the cluster electron counts [18-20]. Among them, the different properties of $GaV_4S_8$ and $GeV_4S_8$ are remarkable. By adding only one electron to the $V_4$ cluster, the magnetic ordering changes from ferro- to antiferromagnetic and the preceding structural distortion of $GaV_4S_8$ leads to orthorhombic symmetry (space group *Imm*2) [11, 12] instead of rhombohedral as in $GeV_4S_8$ [9]. In this paper, we report the synthesis, crystal structures and magnetic properties of the solid solution $Ga_{1-x}Ge_xV_4S_8$ in order to study the transition between these different behaviors in more detail.

**Experimental**

*Synthesis*

Powder samples of $Ga_{1-x}Ge_xV_4S_8$ were prepared by two-step synthesis. First, stoichiometric mixtures of gallium pieces (Alfa Aesar, 99.999%), germanium pieces (Aldrich, 99.999%) and vanadium powder (Smart Elements, 99.9%) were heated at 800°C (50 °C/h) in silica ampoules under argon atmospheres, until binary alloys $Ga_{1-x}Ge_xV_4$ were formed (1-3 heating runs). These precursors were then mixed with stoichiometric amounts of sulfur (Aldrich, 99.99%), sealed in silica tubes under argon and heated to 750 °C (50 °C/h) for 12 h. The samples were then ground and heated subsequently until single phase samples were obtained (1-3 heating runs).

*Crystal structure determination*

Powder patterns were recorded on a Huber G670 Guinier imaging plate diffractometer (Cu-$K_{\alpha1}$ radiation, Ge-111 monochromator) equipped with a closed-cycle He-cryostat. Rietveld refinements were performed with the TOPAS package [21] using the fundamental parameters approach as reflection profiles (convolution of appropriate source emission profiles with axial instrument contributions as well as crystallite microstructure effects). In order to describe small peak half width and shape anisotropy effects, the approach of *Le Bail* and *Jouanneaux* [22] was implemented into the TOPAS program and the according parameters were allowed to refine freely.

Preferred orientation of the crystallites was described with spherical harmonics. An empirical $2\theta$-dependent absorption correction for the different absorption lengths of the GUINIER geometry was applied. As the background between 10 and 25 degrees $2\theta$ shows artifacts from the low-temperature configuration of the GUINIER diffractometer, small sections of this range were excluded from the refinements.

*Magnetic measurements*

Magnetic properties of the samples were measured using a SQUID magnetometer (MPMS-XL5, Quantum Design Inc.). Fine ground powder samples were inserted into gelatine capsules of known diamagnetism and fixed in a straw as sample holder. The magnetic susceptibilities of the samples were collected in a temperature range of 1.8 K to 300 K with magnetic flux densities up to 1 Tesla. Magnetization measurements with magnetic flux densities up to 5 T were recorded at different temperatures. The data were corrected for diamagnetic contributions of the capsule, the straw and the sample using diamagnetic increments [23] and analyzed using the Curie-Weiss law, modified by an additional temperature-independent contribution to the susceptibility $\chi_0$:

$$\chi_{mol} = \frac{C}{T - \theta_{cw}} + \chi_0 \qquad C = \mu_0 \frac{N_A \mu_B^2 \mu_{eff}^2}{3 k_B};$$

**Results and discussion**

*Crystal structures at room temperature*

Room temperature x-ray powder patterns of samples with the nominal compositions $Ga_{1-x}Ge_xV_4S_8$ ($\Delta x = 0.125$) could be completely indexed with cubic face centered unit cells. The lattice parameters of $GaV_4S_8$ (966.1 pm) and $GeV_4S_8$ (965.5 pm) differ by only 0.6 pm. This is close to our experimental error, thus we were not able to deduce the composition reliably from the lattice parameters. However, a series of EDX measurements confirmed the nominal compositions within a tolerance of 10% as shown in Fig. 3. Rietveld refinements of all powder patterns were successful by using the structural parameters of $GaV_4S_8$ as initial models; a typical fitted pattern is shown in Fig. 4. No indication of a deviation from full occupation of any site was detected. This was additionally confirmed by x-ray data of a single crystal with the nominal composition $Ga_{0.5}Ge_{0.5}V_4S_8$, which are not presented here. The results of the Rietveld refinements are compiled in Tab. 1.

*Crystal structures at low temperatures*

Both ternary compounds GaV$_4$S$_8$ and GeV$_4$S$_8$ undergo second order structural phase transitions at 38 K and 30 K, respectively, in both cases well above the onset of magnetic ordering at 15 K and 18 K. While GaV$_4$S$_8$ becomes rhombohedral (space group *R3m*), the low temperature phase of GeV$_4$S$_8$ has orthorhombic symmetry (space group *Imm*2). A report by *Chudo* et al. about a rhombohedral (*R3m*) distortion of GeV$_4$S$_8$ was recently disproved [11]. In general, we have observed suppressions of the structural transitions in GaMo$_4$S$_8$-type compounds with mixed metal site occupations like Ga$_x$V$_{4-y}$Cr$_y$S$_8$ [18], GaNb$_{4-x}$Mo$_x$S$_8$ [19] and GaV$_{4-x}$Mo$_x$S$_8$ [24]. Since we do not introduce disorder in the V$_4$-cluster units in the case Ga$_{1-x}$Ge$_x$V$_4$S$_8$, the structural distortions may rather persist.

In order to check for symmetry reduction, we have recorded x-ray powder patterns of all samples at 10 K. Broadening or even splitting of some reflections was observed, best visible for the (440) reflection close to 54°. We have shown earlier [11], that the splitting of this reflection in two components (220) and (208) indicates a rhombohedral distortion (*R3m*), whereas three components (400), (224), and (040) are only consistent with an orthorhombic lattice (*Imm*2). Fig. 5 shows the (440) reflections of the solid solution in more detail. At low germanium contents ($x = 0.125 - 0.375$), we observe two peaks of identical intensity in agreement with the rhombohedral distortion as known from GaV$_4$S$_8$. The profile of Ga$_{0.5}$Ge$_{0.5}$V$_4$S$_8$ has already three components (one central peak with two satellites), which are not yet separated due to the limited instrument resolution. With increasing germanium substitution, the three peaks become more separated and the pattern is finally similar to the low temperature phase of GeV$_4$S$_8$, which is included in Fig. 5 for comparison.

The low temperature diffraction data unequivocally shows structural distortions in all samples of the series Ga$_{1-x}$Ge$_x$V$_4$S$_8$. We observe a gradual transition between rhombohedral distortions known from GaV$_4$S$_8$ to orthorhombic lattices as observed in GeV$_4$S$_8$ with an inflection close to Ga$_{0.5}$Ge$_{0.5}$V$_4$S$_8$. Fig. 6 shows the Rietveld fit of Ga$_{0.5}$Ge$_{0.5}$V$_4$S$_8$ at 10 K. The structural data of the low temperature phase GeV$_4$S$_8$ (space group *Imm*2) [11] has been used as initial parameters and the refinements resulted in very similar structural parameters for Ga$_{0.5}$Ge$_{0.5}$V$_4$S$_8$. As for the pure germanium compound, the distortion is mainly caused by the change of the ideal V$_4$ tetrahedra into a butterfly-like structure, where one V–V bond is shortened by 6.5 pm and the opposing elongated by 2.5 pm. This clearly demonstrates the identical distortion motifs of $Ga_{0.5}Ge_{0.5}V_4S_8$ and $GeV_4S_8$ and we can safely assume the same also for the higher germanium contents ($x > 0.5$).

*Magnetism*

The inverse magnetic susceptibilities of the $Ga_{1-x}Ge_xV_4S_8$ samples are displayed in Fig. 7. The decreasing slopes of the curves above 100 K indicate the gradual filling of the $V_4$ cluster orbitals that yield increased magnetic moments. We find an almost linear dependency of the effective magnetic moments $\mu_{eff}$ on the germanium content as shown in Fig. 8, which are in agreement with those expected from the cluster MO approach. The magnetic data extracted from the Curie-Weiss analysis are collected in Tab. 2. Anomalies in the $\chi^{-1}(T)$ plots appear around 30 K in all cases, but more pronounced in gallium-rich samples. These more or less sharp drops are caused by the second order structural phase transitions. In the course of these, the $V_4$ cluster MO's become reorganized according to the lowered symmetry, *i. e.* space group *R3m* for $x < 0.5$ and space group *Imm*2 for $x \geq 0.5$ as shown in the previous chapter.

Fig. 9 shows the isothermal magnetization curves measured at 1.8 K. We observe a gradual transition from ferromagnetic to antiferromagnetic behavior as the germanium fraction increases. The magnetization of the samples with compositions $0.25 \leq x \leq 0.75$ (insert in Fig. 9) still show typical ferromagnetic saturation and hysteresis, but the moments decrease strongly with *x*. Since we can assume certain (small) inhomogeneities in the Ga/Ge distributions, we suggest that the observed curves are superpositions of Ge-rich antiferromagnetic and Ga-rich ferromagnetic domains, which accumulate to the observed magnetic moments. The magnetization of $Ga_{0.25}Ge_{0.75}V_4S_8$ is still reminiscent of an antiferromagnetic phase with small ferromagnetic contamination, while the curve $Ga_{0.125}Ge_{0.875}V_4S_8$ is finally almost linear as expected for an antiferromagnet.

**Acknowledgements**

This work was financially supported by the Deutsche Forschungsgemeinschaft.

Table 1. Crystallographic data of $Ga_{0.5}Ge_{0.5}V_4S_8$ at 300 K and 10 K.

| Temperature | 300 K | 10 K |
|---|---|---|
| Space group | $F\bar{4}3m$ | $Imm2$ |
| Molar mass (g mol$^{-1}$) | 531.443 | 531.443 |
| Lattice parameters (pm) | $a = 965.24(3)$ | $a = 683.66(2)$, $b = 681.25(2)$, $c = 964.54(1)$ |
| Cell volume, (nm$^3$) | 0.89923(8) | 0.44923(2) |
| Density, (g cm$^{-3}$) | 3.93 | 3.93 |
| $\mu$ (mm$^{-1}$) | 54.2 | 54.2 |
| Z | 4 | 2 |
| Data points | 17201 | 16000 |
| Reflections | 40 | 153 |
| $d$ range | 1.005 – 6.320 | 1.005 – 6.320 |
| Excluded $2\theta$ range(s) | - | 20-25.5 |
| Constraints | 1 | 6 |
| Atomic variables | 7 | 16 |
| Profile variables | 6 | 6 |
| Anisotropy variables | 12 | 18 |
| Background variables | 48 | 30 |
| Other variables | 5 | 12 |
| $R_P$, $wR_P$ | 0.007, 0.011 | 0.017, 0.023 |
| $R_{bragg}$, $\chi^2$ | 0.006, 1.634 | 0.005, 1.134 |

**Atomic parameters:**

| | | | | |
|---|---|---|---|---|
| Ga/Ge | $4a$ (0, 0, 0); $U_{iso} = 82(1)$ | Ga/Ge | $2b$ (0, 0, z); $z = -0.0078$; $U_{iso} = 84(5$ | |
| V | $16e$ (x, x, x) | V1 | $4c$ (x, 0, z) | |
| | $x = 0.60495(2)$; $U_{iso} = 89(2)$ | | $x = 0.2181(7)$; $z = 0.3845(4)$; $U_{iso} = 89(3)$ | |
| | | V2 | $4d$ (0, y, z) | |
| | | | $y = 0.7944(3)$; $z = 0.5936(5)$; $U_{iso} = 89(3)$ | |
| S1 | $16e$ (x, x, x) | S11 | $4c$ (x, 0, z) | |
| | $x = 0.36959(4)$; $U_{iso} = 116(3)$ | | $x = 0.2600(9)$; $z = 0.6163(5)$; $U_{iso} = 101(3)$ | |
| | | S12 | $4d$ (0, y, z) | |
| | | | $y = 0.7443(5)$; $z = 0.3539(5)$; $U_{iso} = 101(3)$ | |
| S2 | $16e$ (x, x, x) | S21 | $4c$ (x, 0, z) | |
| | $x = 0.86486(4)$; $U_{iso} = 28(3)$ | | $x = 0.2753(8)$; $z = 0.1189(5)$; $U_{iso} = 101(3)$ | |
| | | S22 | $4d$ (0, y, z) | |
| | | | $y = 0.7289(5)$; $z = 0.8536(6)$; $U_{iso} = 101(3)$ | |

**Selected bond lengths (pm):**

| | | | |
|---|---|---|---|
| Ga/Ge – S2 | 225.9(1) ×4 | Ga/Ge – S21 | 224.4(6)×2 |
| | | Ga/Ge – S22 | 228.0(5)×2 |
| V – S1 | 229.8(1)×3 | V1 – S11 | 225.4(7)×1 |
| | | V1 – S12 | 231.2(5)×2 |
| | | V2 – S11 | 227.3(5)×2 |
| | | V2 – S12 | 233.7(7)×1 |
| V – S2 | 254.2(1) ×3 | V1 – S21 | 259.2(7)×1 |
| | | V1 – S22 | 249.7(5)×2 |
| | | V2 – S21 | 253.8(4)×2 |
| | | V2 – S22 | 254.8(7)×2 |
| V – V | 286.5(1) ×3 | V1 – V2 | 298.2(10)×1 |
| | | V1 – V1 | 287.3(5)×1 |
| | | V2 – V2 | 280.1(5)×1 |

Table 2. Magnetic data of $Ga_{1-x}Ge_xV_4S_8$

| $x$ nominal | $\mu_{eff}$ ($\mu_B$) | $\theta_{CW}$ (K) | $\chi_0$ (mol$^{-1}$m$^3$) |
|---|---|---|---|
| 0 | 1.52 | −17 | $1.2 \times 10^{-8}$ |
| 0.125 | 1.66 | −28 | $1.2 \times 10^{-8}$ |
| 0.250 | 1.82 | −58 | $1.1 \times 10^{-8}$ |
| 0.375 | 1.94 | −67 | $1.0 \times 10^{-8}$ |
| 0.500 | 2.22 | −54 | $9.2 \times 10^{-9}$ |
| 0.625 | 2.25 | −63 | $9.4 \times 10^{-9}$ |
| 0.750 | 2.48 | −44 | $8.5 \times 10^{-9}$ |
| 0.875 | 2.69 | −54 | $8.3 \times 10^{-9}$ |
| 1 | 2.80 | −40 | $1.1 \times 10^{-8}$ |

**Figures and Captions**

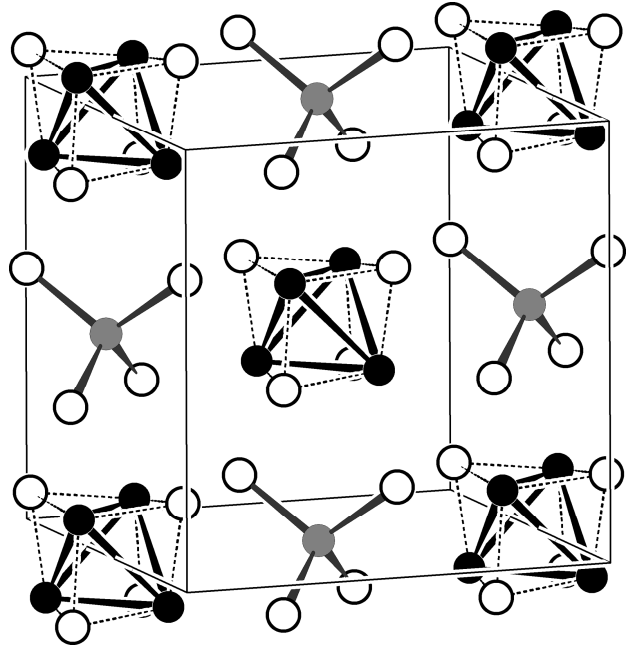

Figure 1. Crystal structure of $GaV_4S_8$ (V black, Ga grey, S white)

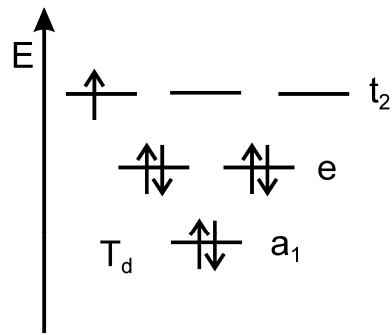

Figure 2. Cluster MO scheme of cubic $GaV_4S_8$

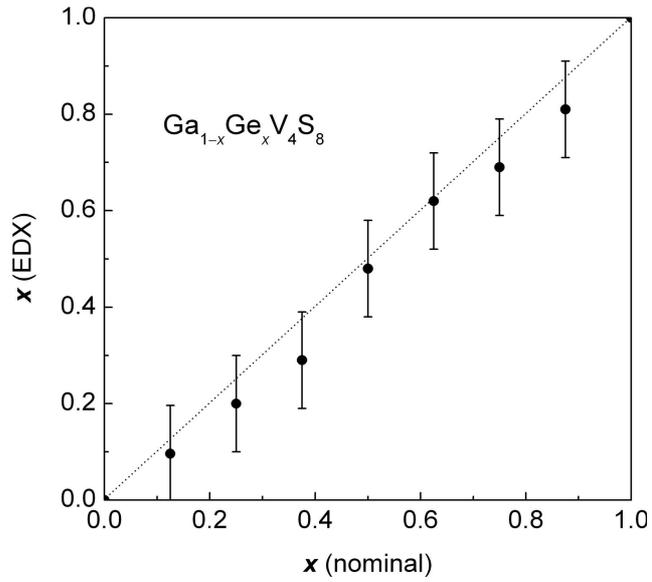

Figure 3. Germanium contents of $Ga_{1-x}Ge_xV_4S_8$ obtained from EDX measurements versus the nominal compositions.

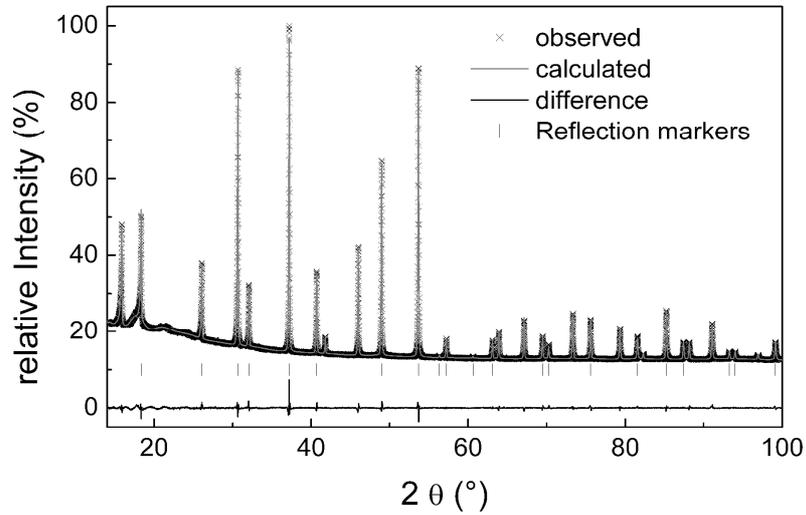

Figure 4. X-ray powder pattern and Rietveld-Fit of $Ga_{0.5}Ge_{0.5}V_4S_8$ measured at room temperature.

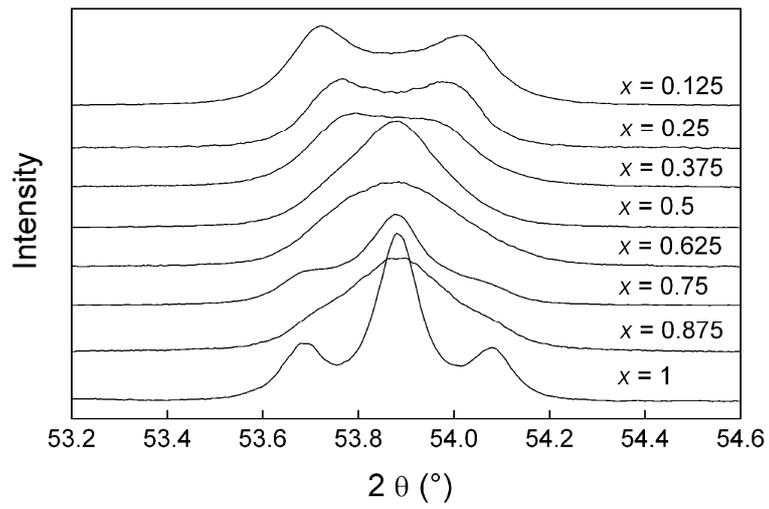

Figure 5. Sections around the (440) reflections of $Ga_{1-x}Ge_xV_4S_8$ samples measured at 10 K

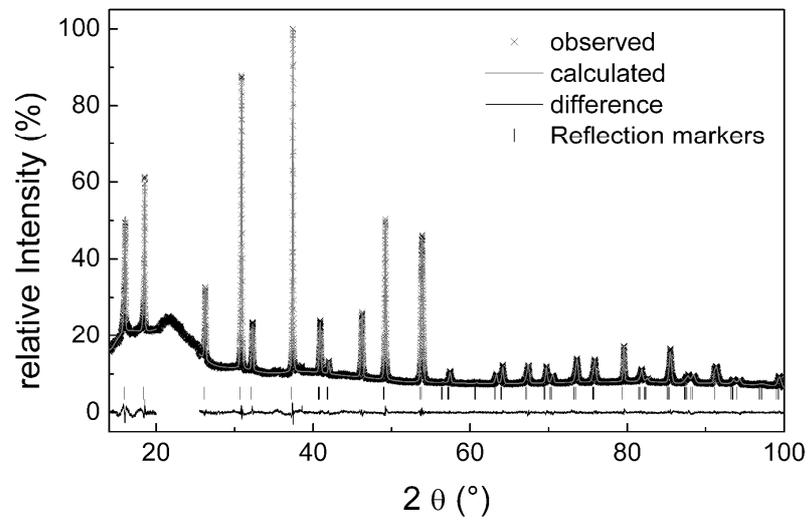

Figure 6. X-ray powder pattern and Rietveld-Fit of $Ga_{0.5}Ge_{0.5}V_4S_8$ measured at 10 K.

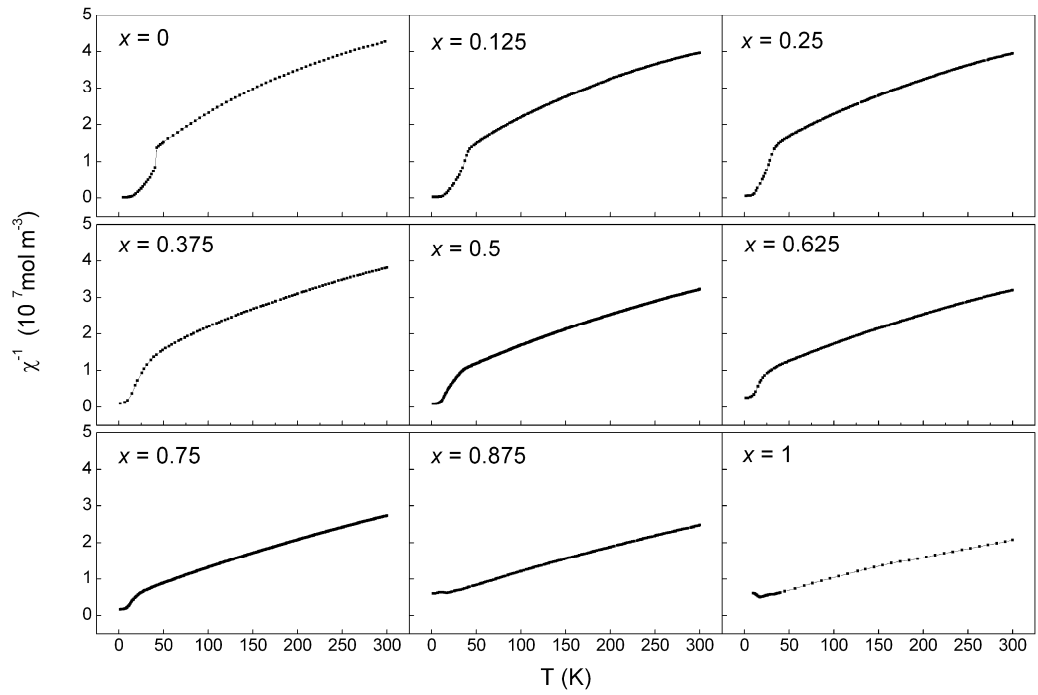

Figure 7. Inverse magnetic susceptibility of $Ga_{1-x}Ge_xV_4S_8$ samples.

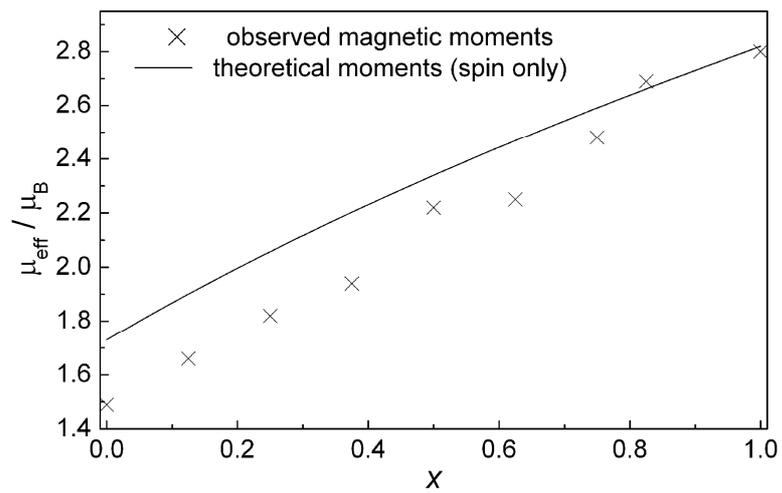

Figure 8. Effective magnetic moments of $Ga_{1-x}Ge_xV_4S_8$

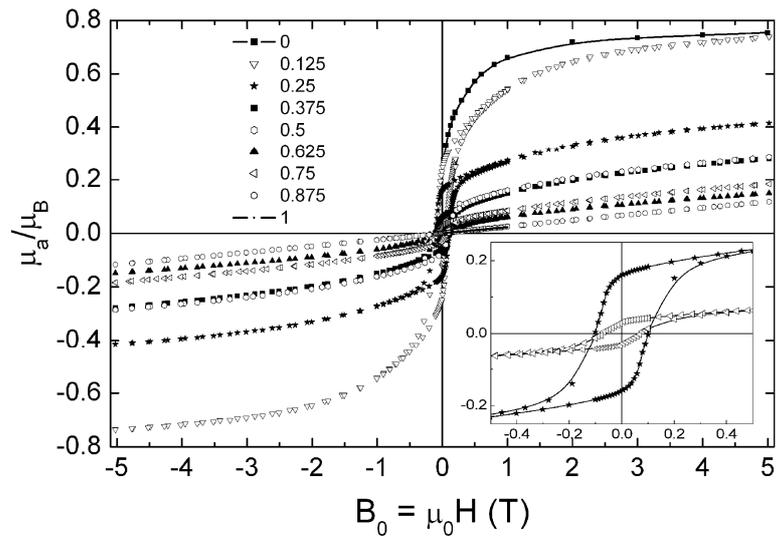

Figure 9. Isothermal magnetization of $Ga_{1-x}Ge_xV_4S_8$ samples.